\documentclass{article}

\usepackage{arxiv}
\usepackage{cite}
\usepackage{amsmath,amssymb,amsfonts}
\usepackage{algorithmic}
\usepackage{graphicx}
\usepackage{textcomp}
\usepackage{mathrsfs}

\date{}
\title{\LARGE \bf
Dimensionality Reduction with Koopman Generalized Eigenfunctions
}

\author{Simone Martini, Margareta Stefanovic and Kimon P. Valavanis
\thanks{Simone Martini is now at the University of Cincinncati. This work is the result of the Simone Martini postdoctoral work at the University of Denver. The authors are from the Department of Electrical and Computer Engineering at the Ritchie School of Engineering and Computer Science, University of Denver, USA,
{\tt\small simone.martini@du.edu, marti6so@ucmail.uc.edu},
{\tt\small margareta.stefanovic@du.edu}, 
{\tt\small kimon.valavanis@du.edu}}
}

\begin{document}

\maketitle
\thispagestyle{empty}
\pagestyle{empty}

\begin{abstract}

This paper presents a methodology to achieve lower-dimensional Koopman quasi-linear representations of nonlinear system dynamics using Koopman generalized eigenfunctions. The proposed approach considers the analytically derived Koopman formulation of rigid body dynamics, but it can be extended to any data-driven or analytically derived generalized eigenfunction set. It achieves a representation for which the number of Koopman observables matches the number of inputs allowing for Koopman linearization control solutions rather than resorting to the least squares approximation method adopted in high dimensional Koopman formulations. Through a linear combination  of Koopman generalized eigenfunctions a new set of Koopman generalized eigenfunction is constructed so that the zero order truncation approximate a Koopman eigenfunction which can be used to design linear control strategies to steer the dynamics of the original nonlinear system. The proposed methodology is tested by designing a linear quadratic (LQ) flight controller for a quadrotor UAV. Numerical and Hardware-in-the-loop (HIL) simulations validate the applicability and real-time implementability of the proposed approach in the presence of noise and sensor delays. The main advantage of the proposed method is the realization of a fully actuated Koopman based model which, in the case of the underactuated quadrotor system, allows to achieve trajectory tracking through a single linear control loop.
\end{abstract}

\section{Introduction}
This paper introduces a methodology to obtain a lower dimensional Koopman quasi-linear representation (characterized by a linear state matrix and a state dependent nonlinear control matrix) of nonlinear dynamics using Koopman generalized eigenfunctions.

Koopman operator theory \cite{koopman1931hamiltonian} serves as a link between linear and nonlinear system theory, establishing itself as a promising general framework for analyzing nonlinear systems dynamics. Its resurgence is partly attributed to advancements in dynamic mode decomposition (DMD) \cite{schmid2010dynamic} , which has facilitated the modeling and analysis of complex nonlinear systems, such as fluid flows, through data-driven approaches \cite{rowley2009spectral}. 

The Koopman operator is an infinite dimensional linear operator that maps the evolution of the state observables (i.e. nonlinear functions of the state variables) \cite{kaiser2021data,Bevanda,Mezic}, embedding the original state nonlinear dynamics into Koopman observables' linear dynamics. Recent applications that have employed the Koopman framework have been used to model and control systems in aerospace
\cite{Abraham2, mamakoukas2022robust,folkestad2022koopman,manaa2024koopman} and robotics \cite{sinha2022koopman,goyal2022impedance,Bruder}. Additionally, Koopman-based analytical derivation of position and attitude dynamics has been the focus of research \cite{Chen,Zinage,Zinage2}, and \cite{martini2023koopman,martini2024koopman,martini2024koopman_rigid}.

A key limitation of Koopman based modeling is the resulting high-dimensionality of the Koopman operator, which can render it impractical for control design. A general solution to this issue is to adopt a least square approximation which allows to invert the control matrix and design linear control strategies in the Koopman based coordinates. However, this approach is strictly dependent on the approximation error and, in the authors' experience, requires an extensive controller tuning effort \cite{martini2024koopman_rigid}. A possible solution to mitigate the high dimensionality is to employ a special class of Koopman observables named Koopman eigenfunctions and Koopman generalized eigenfunctions which are associated with the principal system dynamics (since their span include any observable function, comprising the state) and therefore allow for a better approximation of the original system dynamics in a lower dimensional space \cite{kaiser2021data,9022864}.

This work poses the focus on Koopman formulations based on the Koopman generalized eigenfunctions. Starting from the relation between  Koopman eigenfunctions and linear systems eigenmodes, highlighted in \cite{rowley2009spectral}, we propose a methodology to design control strategies in terms of approximated Koopman eigenfunctions constructed from a linear combination of Koopman generalized eigenfunctions. 
Although implemented on the analytically derived Koopman formulation of rigid body dynamics presented by the authors in \cite{martini2024koopman_rigid}, the methodology can be extended to any data-driven or analytically derived generalized eigenfunction set. The presented approach achieves a representation for which the number of Koopman observables matches the number of inputs allowing exact inversion of the control matrix, avoiding the least square approximation error and simplifying controller tuning. Application to the Koopman modeling formulation of a quadrotor unmanned aerial vehicle (UAV) leads to a fully actuated quasi-linear model which is used for the design of a linear quadratic (LQ) flight controller. Hardware in the loop (HIL) simulations in presence of noise and sensor delays validate the real-time implementability of the approach.

The remaining of the paper is organized as follow: Section \ref{sec:background} introduces notation and background information about Koopman and the Koopman formulation of rigid body dynamics. Section \ref{sec:method} outline the control methodology based on the proposed Koopman dimensionality reduction. Using the presented approach, the Koopman lower dimensional model of rigid body dynamics is derived in Section \ref{sec:Kooplow} and is then employed for the control design of an underactuated quadrotor UAV in Section \ref{sec:quadrotor}. Finally, conclusion and future works are discussed in \ref{sec:conclusions}. 

\section{Notation \& Background Information}\label{sec:background}
\subsection{Notation}
Denote $\mathbf{0_n}$ and $I_n$ as the $n \times n$ zero matrix and identity matrix, respectively. Fix the unit vector in the third dimension as $\mathbf{e_3}= [0,0,1]^{\top}$. Given the vectors $a, b \in \mathbb{R}^{n}$ and matrices $A\in\mathbb{R}^{n\times n},B\in\mathbb{R}^{m\times m}$, let the operator ${\mathcal{T}}$ refer to the block transpose operation such that $[a,b]^{\mathcal{T}} \triangleq [a^{\top},b^{\top}]^{\top}$, define the operation $diag$ so that $diag(a) = a^{\top}I_n$, and denote the matrix $C=blkdiag(A,B)\in\mathbb{R}^{(n+m)\times(n+m)}$ as the block diagonal matrix with $A$ and $B$ diagonal elements. Let $S(\cdot)\in \mathbb{R}^{3\times 3}$ denote the skew symmetric matrix which can be used interchangeably to represent the cross product since, given $a, b \in \mathbb{R}^{3}$, $S(a)b \triangleq a \times b$.

\subsection{Koopman Theory Fundamentals}

Consider the general continuous-time autonomous nonlinear dynamics on a smooth $n$-dimensional manifold $\mathcal{M}$, where the evolution of the state vector $x \in \mathcal{M}\subset\mathbb{R}^n$ is represented by
$\frac{d}{{dt}}x(t) = f(x)$
with the corresponding smooth Lipschitz continuous flow $F_t:\mathcal{M}\mapsto\mathcal{M}$, such that 
$F_t(x({t_0})) = x({t_0}) + \int_{{t_0}}^{{t_0} + t} {f(x(\tau ))} d\tau$,
with initial time $t_0\geq0$. The Koopman operator semigroup $\mathcal{K}_{t\geq0}$, referred to as the Koopman operator, acting on scalar function $b(x):\mathcal{M}\mapsto\mathbb{C}$, called observable, is defined on the state space $\mathcal{M}$ as
$\mathcal{K}_tb=b\circ F_t$. The Koopman operator is, therefore, an infinite dimensional linear operator that maps the evolution of the state observables (i.e. nonlinear functions of the state variables) \cite{kaiser2021data,Bevanda,Mezic,mezic2022numerical}. 
A Koopman eigenfunction $\varphi$, with eigenvalue $\lambda\in\mathbb{C}$, is an observable such that 
\begin{equation}
    \frac{d}{{dt}}\varphi(F_t(x))=\lambda\varphi((F_t(x)))
\end{equation}
Additionally, considering the eigenvalue $\lambda$ and the vector of Koopman observables $\varkappa = \left[\varkappa_1,\cdots,\varkappa_{n,\lambda}\right]^{\top}$, for which
\begin{equation}\label{eq:geneig}
    \frac{d}{{dt}}\varkappa(F_t(x)) = {J_{\lambda}} \varkappa(F_t(x))
\end{equation}
with $J_{\lambda}$ being the Jordan block 
\begin{equation}
J_{\lambda} = 
\begin{bmatrix}
\lambda & 1 & \\
 & \ddots & 1 \\
 &  & \lambda
\end{bmatrix}
\end{equation}
it follows that the $\mathrm{span}\left\{\varkappa_1,\cdots,\varkappa_{n,\lambda}\right\}$ is a Koopman invariant subspace and $\varkappa$ is the vector of Koopman generalized eigenfunctions \cite{9022864} which can be used to embed the original nonlinear system dynamics.

\subsection{Analytically Derived Koopman Formulation of Rigid Body Dynamics}\label{sec:2c}
The attitude and position rigid body dynamics are presented as a set of nonlinear differential equations \cite{siciliano2010robotics}:
\begin{subequations}
    \begin{align}
\dot R &= RS(\nu) \\ J\dot \nu &= M - S(\nu)J\nu \\ \dot p &= Rv \\ \dot v &= \frac{1}{m}F - S(\nu)v - gR^{\top}\mathbf{e_3}
\end{align}\label{eq:eqmot}
\end{subequations}

where $R\in \mathbb{R}^{3\times 3}$ is the rotation matrix based on the Euler angles configuration $\eta \in \mathbb{R}^{3}$; $\nu, v \in \mathbb{R}^{3}$ are the angular and linear velocity expressed in the body fixed frame; $J = diag(I_x,I_y,I_z)\in \mathbb{R}^{3\times 3}$ is the inertia tensor; $M, F \in \mathbb{R}^{3}$ are the external torque and force; $p \in \mathbb{R}^{3}$ is the position vector, $m$ is the total mass, and $g$ is the gravitational acceleration. The rigid body dynamics state vector is defined as $x = \left[p,\eta,\dot p, \nu\right]^{\top}$.

In \cite{martini2024koopman_rigid}, a Koopman analytical formulation of rigid body dynamics is proposed. 
This is achieved by analytically deriving a special set of Koopman observables known as generalized eigenfunction of the Koopman operator $\varkappa$ \cite{9022864}. For completeness, a summary of the formulation is reported here. The analytically derived Koopman observable set for the rigid body dynamics is 
\begin{equation}
   \mathbf{x}(x) = (\nu_0, \{\nu_k\}^{N_{\nu}-1}_{k=1}, z_0, \{z_k\}^{N_z-1}_{k=1}) \in \mathbb{R}^{N \times 1}
\end{equation}
where $N_{\nu}$ and $N_z$ are the dimension of the position and attitude observable set with components $\nu_k$ and $z_k$, respectively and \( N = 3N_{\nu} + 3N_z \) is the complete dimension of the finite truncation of the Koopman formulation. The system dynamics are described by
\begin{equation}
    \dot{\mathbf{x}} = A\mathbf{x} + B(\mathbf{x})u
\end{equation}

with \( u = \left[M,F \right]^{\mathcal{T}} \in \mathbb{R}^{6 \times 1} \) as the input vector, and \( A \in \mathbb{R}^{N \times N} \), \( B(\mathbf{x}) \in \mathbb{R}^{N \times 6} \) representing the state matrix and state-dependent control matrix, respectively. The Koopman observables \( \nu_k \) and \( z_k \) are defined as:
\begin{subequations}
    \begin{equation}
        \nu_k = J^{-1} \sum_{n=0}^{k} \binom{k}{n} S(\gamma_n) \nu_{k-n}
    \end{equation}
    \begin{equation}
        z_k = p_k + \alpha_k v_{k-1} + \beta_k g_{k-2}
    \end{equation}
\end{subequations}

with \( \nu_0 = \nu \), \( \gamma_0 = J\nu \), \( \alpha_k = k \), and \( \beta_k = \alpha_{k-1} + \beta_{k-1} \). Additionally, the observables of position \( p_k \), velocity \( v_k \), and gravity vector \( g_k \) are computed through recursive differentiation using the following relations:
\begin{subequations}
    \begin{equation}
        p_{k+1} = \sum_{n=0}^{k} \binom{k}{n} S^{\top}(\nu_n) p_{k-n},
    \end{equation}
    \begin{equation}
        v_{k+1} = \sum_{n=0}^{k} \binom{k}{n} S^{\top}(\nu_n) v_{k-n},
    \end{equation}
    \begin{equation}
        g_{k+1} = \sum_{n=0}^{k} \binom{k}{n} S^{\top}(\nu_n) g_{k-n}.
    \end{equation}
\end{subequations}

The quantities \( g_0 = R^{\top} g \mathbf{e_3} \), \( v_0 = R^{\top} \dot{p} \), and \( p_0 = R^{\top} p \) define the zero order observables.

By applying an appropriate linear transformation matrix $V_J$, the system can be transformed into Jordan form, so that the resulting observables \( \varkappa(x) = V_J\mathbf{x}(x) \) are composed of the Koopman generalized eigenfunctions sets. The state dynamics are then given by
\begin{equation}
\dot{\varkappa} = A_J \varkappa + B_J(\varkappa)u \label{eq:completeKsys}
\end{equation}

where \( A_J \) consists of six Jordan blocks, effectively organizing the vector \( \varkappa \) into six sets of Koopman operator generalized eigenfunctions $ \varkappa = \left( \varkappa_1, \varkappa_2, \varkappa_3, \varkappa_4,\varkappa_5,\varkappa_6 \right)$. Three of these blocks correspond to the attitude dynamics (\(\nu_0\)) and the other three to the ``rotated position" \cite{martini2024koopman_rigid} dynamics (\( p_0 \)).

To design a control strategy, the system is rewritten as $\dot \varkappa = A\varkappa + B^{\star}U^{\star}(\varkappa)$ where $U^{\star}(\varkappa) = B(\varkappa)u$ and $B^{\star} = I_N$. Hence, the control law can be formulated for a linear time invariant (LTI) system. The control input is then computed by solving the least square optimization problem:
\begin{equation}
    \{\text{min:} (B_J(\varkappa)u-B_J^{\star}U^{\star}(\varkappa))^{\top}(B_J(\varkappa)u-B_J^{\star}U^{\star}(\varkappa))\}
\end{equation}
whose solution is $u = B_J^{\dagger}(\varkappa)B^{\star}U^{\star}(\varkappa)$. To guarantee the validity of such procedure it is crucial that the optimization error $B_J(\varkappa)u-B_J^{\star}U^{\star}(\varkappa)$ remains sufficiently small during operation. Since this is not always feasible, the methodology presented in this work aims at matching the number of inputs to the number of outputs to obtain a nonsingular control matrix for which $B_J^{\dagger}(\varkappa)=B_J^{-1}(\varkappa)$.

\section{Proposed Control Methodology}\label{sec:method}
The proposed methodology is the outgrowth of the authors' note provided in Appendix \ref{appendix} but shifts the focus on Koopman based representation of nonlinear dynamics. Although applied to the Jordan form of the analytically derived Koopman formulation of rigid body dynamics, the approach may be extended to any Koopman generalized eigenfunctions set, either data-driven or analytically derived. The approach exploits the relation of Koopman eigenfunctions and linear systems eigenmodes, highlighted in \cite{rowley2009spectral}, and defines a control strategy in terms of a linear combination of Koopman generalized eigenfunctions designed to approximate a Koopman eigenfunction and therefore reducing the system dimensionality.

\subsection{Third Order SIMO Case}
For simplicity purposes, a third order linear system is initially considered 
\begin{equation}
    \dot{\mathbf{x}} = A \mathbf{x} + Bu \label{eq:linearsys3}
\end{equation}
with
\begin{align}
    \mathbf{x} = \begin{bmatrix}
        \mathbf{x_0} \\ \mathbf{x_1} \\ \mathbf{x_{2}}
    \end{bmatrix},
    \quad A = \begin{bmatrix}
        0 & 1 & 0\\
        0 & 0 & 1\\
        0 & 0 & 0\\
    \end{bmatrix},
    \quad B = \begin{bmatrix}
        b_0 \\ b_1 \\ b_{2}
    \end{bmatrix}
\end{align}
The system is unstable due to the repeated zero eigenvalue with multiplicity three.  Since $A$ is a Jordan block with zero eigenvalue, $\mathbf{x}$ is equivalent to a vector of Koopman generalized eigenfunctions. It is desirable to design a controller $K$ which allows to place the eigenvalues of the closed-loop system $A-BK$ in the left hand plane. To this end, consider the desired characteristic polynomial
\begin{equation}
    (s - \lambda^*)(s - \lambda_1)(s - \lambda_2) = 0 \label{eq:characteristic}
\end{equation}
From the approach presented in Appendix \ref{appendix}, \eqref{eq:characteristic} can be rearranged as
\begin{subequations}
    \begin{align}
    &s(s - \lambda_1)(s - \lambda_2) = \lambda^*(s - \lambda_1)(s - \lambda_2)\\
    &s\left( \beta_2s^2 +\beta_1s + \beta_0 \right) = \lambda^* \left( \beta_2s^2 +\beta_1s + \beta_0 \right) \\
    &\overset{\mathcal{L}}{\rightarrow} \frac{d}{dt}{\left( \beta_2\mathbf{x_2} + \beta_1\mathbf{x_1} +  \beta_0 \mathbf{x_0}  \right)} =  \lambda^* \left( \beta_2\mathbf{x_2} + \beta_1\mathbf{x_1} +  \beta_0 \mathbf{x_0}  \right) \label{eq:eigenfunc}
\end{align}
\end{subequations}
where $\beta_0 = \lambda_1 \lambda_2$, $\beta_1 = - (\lambda_1 + \lambda_2)$, and $\beta_0 = 1$. In other words, $\tilde{\varphi}^*(\mathbf{x}) = \beta_2\mathbf{x_2} + \beta_1\mathbf{x_1} +  \beta_0 \mathbf{x_0}$ acts as a Koopman operator eigenfunction for the closed-loop system and is obtained through a linear combination of Koopman generalized eigenfunctions $\mathbf{x}$. Note that the coefficients $\beta_i$ are ones of a Hurwitz polynomial and so they all have same sign. From Appendix \ref{appendix}, it is possible to design a state feedback control input such that \eqref{eq:eigenfunc} is satisfied and $ \tilde{\varphi}^* (\mathbf{x})$ is a Koopman operator eigenfunction, however, the approach proposed in this work aims at designing a control input in terms of the closed loop desired scalar eigenfunction (instead of the original state vector), which further simplifies controller design
\begin{equation}
    u = -k\tilde{\varphi}^*(\mathbf{x}) \label{eq:control_phi}
\end{equation}
Considering system \eqref{eq:linearsys3}, the evolution of \eqref{eq:eigenfunc} is 
\begin{align}
    &\frac{d}{dt}{\left( \beta_2\mathbf{x_2} + \beta_1\mathbf{x_1} +  \beta_0 \mathbf{x_0}  \right)} = \nonumber\\
    &\qquad\beta_1 \mathbf{x_2} + \beta_0 \mathbf{x_1}  + \underbrace{\left(\beta_2 b_2 +\beta_1 b_1 + \beta_0 b_0 \right)}_{\tilde{b}^*}u\label{eq:boh}
\end{align}
which approximates the evolution of an eigenfunction with zero eigenvalue for 
\begin{equation}
    \beta_1 \mathbf{x_2} +\beta_0 \mathbf{x_1} \ll \tilde{b}^*u \label{eq:condition}
\end{equation}
so that 
\begin{equation}
    \frac{d}{dt} \tilde{\varphi}^* (\mathbf{x}) \approx \tilde{b}^* u = 0\cdot\tilde{\varphi}^* + \tilde{b}^* u
\end{equation}
Applying \eqref{eq:control_phi} to \eqref{eq:boh}, we obtain
\begin{align}
    &\frac{d}{dt} \tilde{\varphi}^* (\mathbf{x}) = \beta_1 \mathbf{x_2} +\beta_0 \mathbf{x_1}  - \tilde{b}^* k\tilde{\varphi}^* (\mathbf{x}) \label{eq:poleplace1}
\end{align}
without condition \eqref{eq:condition}, it is clear that designing the control using $\tilde{\varphi}^* (\mathbf{x})$ violates \eqref{eq:eigenfunc}, and $\tilde{\varphi}^* (\mathbf{x})$ is not a Koopman eigenfunction. However, selecting a control input which satisfies \eqref{eq:condition} 
\begin{equation}
    k = -\frac{\lambda^*}{\tilde{b}^*}
\end{equation}
with $\lambda^*, \beta_0, \beta_1$, and $\beta_2$ such that $\beta_1 \mathbf{x_2} +\beta_0 \mathbf{x_1} \ll \tilde{b}^* k\tilde{\varphi}^* (\mathbf{x})$ allows to approximate
\begin{align}
    &\frac{d}{dt} \tilde{\varphi}^* (\mathbf{x}) \approx - \tilde{b}^* k\tilde{\varphi}^* (\mathbf{x}) = \lambda^* \tilde{\varphi}^* (\mathbf{x}) \label{eq:approx1}
\end{align}
Hence, by proper selection of $\beta_i$, a controller is designed so that a the linear combination of Koopman generalized eigenfunctions approximates a Koopman eigenfunction with eigenvalue $\lambda^*$.

\subsection{$n$-th Order SIMO Case}
Consider now the general high dimensional single input multi output (SIMO) case
\begin{equation}
    \dot{\mathbf{x}} = A \mathbf{x} + Bu \label{eq:lineargeneralsys}
\end{equation}
with
\begin{align}
    \mathbf{x} = \begin{bmatrix}
        \mathbf{x}_0 \\ \mathbf{x}_1 \\ \vdots \\ \mathbf{x}_{{n-1}}
    \end{bmatrix},
    \quad A = \begin{bmatrix}
        0 & 1  & \cdots & 0\\
        0 & 0 & \ddots & \vdots\\
        0 & 0 & \cdots & 1\\
        0 & 0 & \cdots & 0\\
    \end{bmatrix},
    \quad B = \begin{bmatrix}
        b_0 \\ b_1 \\ \vdots \\ b_{n-1}
    \end{bmatrix}
\end{align}
where $\mathbf{x} \in \mathbb{R}^{n\times1}$, $u \in \mathbb{R}$, $A \in \mathbb{R}^{n\times n}$, $B \in \mathbb{R}^{n\times 1}$. The system is characterized by a $n$-time repeated zero eigenvalue. The above approach is generalized as 
\begin{subequations}
    \begin{align}
    &s\prod_{i=1}^{n-1}(s - \lambda_i) = \lambda^*\prod_{i=1}^{n-1}(s - \lambda_i)\\
    &\overset{\mathcal{L}}{\rightarrow} \frac{d}{dt}{\underbrace{\left( \beta^{\top}\mathbf{x}\right)}_{\tilde{\varphi}^*(\mathbf{x})} } =  \lambda^* \left( \beta^{\top}\mathbf{x} \right) \label{eq:eigenfunc1}
\end{align}
\end{subequations}
where $\beta = \left[\beta_0, \beta_1, \dots, \beta_{n-1}\right]^{\top} \in \mathbb{R}^{n\times 1}$. Leading to
\begin{align}
    &\frac{d}{dt} \tilde{\varphi}^* (\mathbf{x}) = \beta^{\top} A \mathbf{x} - \beta^{\top}B u \label{eq:poleplace1}
\end{align}
with condition 
\begin{equation}
    \beta^{\top} A \mathbf{x} \ll \beta^{\top}Bu \label{eq:condition1}
\end{equation}
Designing the control input analogously to \eqref{eq:control_phi} with
\begin{equation}
    k = -\frac{\lambda^*}{\beta^{\top}B}
\end{equation}
with $\lambda^*$, and $\beta$ such that $\beta^{\top} A \mathbf{x} \ll \beta^{\top}B k\tilde{\varphi}^*$, allows to approximate
\begin{align}
    &\frac{d}{dt} \tilde{\varphi}^* (\mathbf{x}) \approx - \beta^{\top}B k\tilde{\varphi}^* (\mathbf{x}) = \lambda^* \tilde{\varphi}^* (\mathbf{x}) \label{eq:approx2}
\end{align}
As previously stated, $\tilde{\varphi}^* (\mathbf{x})$ is a linear combination of Koopman generalized eigenfunctions. 
Lastly, the resulting controller $K$ of the original closed-loop system is 
\begin{equation}
    u = -K\mathbf{x} = -k\beta^{\top}\mathbf{x} = -k\tilde{\varphi}^*(\mathbf{x}).
\end{equation}

Highlighting the property for which a linear combination of a linear operator generalized eigenfunctions is itself a generalized eigenfunction of said linear operator, an equivalent result to the above is reached by constructing a new set of Koopman generalized eigenfunctions
\begin{subequations}
\begin{align}
    &\underbrace{\frac{d}{dt}{\left( \beta_{n-1}\mathbf{x_{n-1}} + \dots +  \beta_0 \mathbf{x_0} \right)} }_{\dot{\tilde{\varphi}}^{\{0\}}} = \nonumber\\ &\underbrace{\left( \beta_{n-2}\mathbf{x_{n-1}} + \dots +  \beta_0 \mathbf{x_1} \right)}_{{\tilde{\varphi}^{\{1\}}}} + \underbrace{\left(\beta_{n-1} b_{n-1} +\dots + \beta_0 b_0 \right)}_{\tilde{b}^{\{0\}}}u \\
    &\underbrace{\frac{d}{dt}{\left( \beta_{n-2}\mathbf{x_{n-1}} + \dots +  \beta_0 \mathbf{x_1} \right)} }_{\dot{\tilde{\varphi}}^{\{1\}}} = \nonumber\\ &\underbrace{\left( \beta_{n-3}\mathbf{x_{n-1}} + \dots +  \beta_0 \mathbf{x_2} \right)}_{{\tilde{\varphi}^{\{2\}}}} + \underbrace{\left(\beta_{n-2} b_{n-1} +\dots + \beta_0 b_1 \right)}_{\tilde{b}^{\{1\}}}u \\
        &\qquad \qquad \quad \vdots \nonumber
\end{align}
\end{subequations}
or, in general,
\begin{equation}
    \dot{\tilde{\varphi}}^{\{i\}} = \beta^{\top} A^{i+1} \mathbf{x} + \beta^{\top}A^{i}Bu
\end{equation}
where $i$ is the order of differentiation of the Koopman generalized eigenfunction.
Note that A is a Jordan block and so $A^n = \mathbf{0_{n\times n}}$.
The new system of Koopman generalized eigenfunction is
\begin{equation}
    \dot{\tilde{\varphi}} = A \tilde{\varphi} + \tilde{B}u \label{eq:linearsysnewinf}
\end{equation}
where $\tilde{\varphi} = \left[\tilde{\varphi}^{\{0\}},\dots, \tilde{\varphi}^{\{n-1\}}\right]^{\top}$ and $\tilde{B} = \left[\tilde{b}^{\{0\}},\dots, \tilde{b}^{\{n-1\}}\right]^{\mathcal{T}}$. It is evident that $\tilde{\varphi}^{\{0\}} = \tilde{\varphi}^*$ and applying the approximation of \eqref{eq:approx2} is equivalent to truncate system \eqref{eq:linearsysnewinf} to the zero order.

\subsection{MIMO Case with Matching Number of Inputs and Jordan Blocks}\label{sec:3c}
Consider the multi input multi output (MIMO) system
\begin{equation}
    \dot{\mathbf{x}} = A \mathbf{x} + Bu  \label{eq:MIMOsys}
\end{equation}
with
\begin{align}
    \mathbf{x} = \begin{bmatrix}
        \mathbf{x_{\mathcal{J},1}} \\ \mathbf{x_{\mathcal{J},2}} \\ \vdots \\ \mathbf{x}_{{\mathcal{J},m}}
    \end{bmatrix},
    \; A = \begin{bmatrix}
        {\mathcal{J}_{n_1}} & 0  & \cdots & 0\\
        0 & {\mathcal{J}_{n_2}} & \ddots & \vdots\\
        0 & 0 & \ddots & 0\\
        0 & 0 & \cdots & {\mathcal{J}_{n_\mathbf{m}}}\\
    \end{bmatrix},
    \; B = \begin{bmatrix}
        B_1 \\ B_2 \\ \vdots \\ B_{\mathbf{m}}
    \end{bmatrix}
\end{align}

where  $\mathcal{J}_i$ is the Jordan block matrix with zero eigenvalue of dimension $i$, $\mathbf{x} \in \mathbb{R}^{N\times1}$, $\mathbf{x_{\mathcal{J},i}}\in \mathbb{R}^{n_i\times1}$, $u \in \mathbb{R}^{m\times1}$, $A \in \mathbb{R}^{N\times N}$, $B \in \mathbb{R}^{N\times m}$, $B_i \in \mathbb{R}^{n_i\times m}$, and $N = \sum_{1}^{m}n_i$. System \eqref{eq:MIMOsys} is equivalent to a system composed of same number ($m$) of Koopman generalized eigenfunctions sets as the number of inputs. Considering $\beta^{\top} = blkdiag(\beta_1^{\top},\dots,\beta_{m}^{\top})$ with $\beta_i^{\top} \in \mathbb{R}^{i\times n_i}$, the above approach is adopted to each eigenfunction set
\begin{equation}
        \dot{\tilde{\varphi}}^{\{i\}}_j = \beta^{\top}_j {\mathcal{J}_{n_j}}^{i+1} \mathbf{x_{\mathcal{J},j}} - \underbrace{\beta^{\top}_j{\mathcal{J}_{n_j}}^{i}B_j}_{\tilde{B}^{\{i\}}_j}u
\end{equation}
where $j$ refers to the $j$-th Koopman generalized eigenfunction set and $i$ is the differentiation order. Hence, each newly generated generalized eigenfunction set is defined as
\begin{equation}
        \dot{\tilde{\varphi}}_j =  {\mathcal{J}_{n_j}} {\tilde{\varphi}}_j + \tilde{B}_ju
\end{equation}
where $\tilde{\varphi}_j = \left[\tilde{\varphi}^{\{0\}}_j,\dots, \tilde{\varphi}^{\{n_j-1\}}_j\right]^{\top}$ and $\tilde{B}_j = \left[\tilde{B}^{\{0\}}_j,\dots, \tilde{B}^{\{n_j-1\}}_j\right]^{\mathcal{T}}$. Truncating to the zero order (${\tilde{\varphi}}^{\{0\}}_j = \tilde{\varphi}^{*}_j$ and $\tilde{B}^{\{0\}}_j = \tilde{B}^{*}_j$) leads to $m$ approximated Koopman eigenfunction candidates 
\begin{align}
        \dot{\tilde{\varphi}}^{*} = \beta^{\top} A \mathbf{x}  + \underbrace{\beta^{\top}B}_{B_{\mathbf{low}}}u
\end{align}
where $\dot{\tilde{\varphi}}^{*} = \left[\tilde{\varphi}^{*}_1,\dots, \tilde{\varphi}^{*}_m\right]^{\top}$. Condition \eqref{eq:condition1} is extended to the described MIMO case since $\beta^{\top} \in \mathbb{R}^{m\times N}$, leading to 
\begin{align}
        \dot{\tilde{\varphi}}^{*} \approx B_{\mathbf{low}}u
\end{align}
With the assumption that $B_{\mathbf{low}}$ is invertible, the following control input is design from the extension of the above approach to MIMO
\begin{equation}
    u = -B_{\mathbf{low}}^{-1}u^*= -B_{\mathbf{low}}^{-1}k\tilde{\varphi}^*(\mathbf{x}) \label{eq:control_phiBlowinv}
\end{equation}
where selecting $k = diag(k_1,\dots,k_m)$ and $k_i = -\lambda^*_i$, such that $\beta^{\top} A \mathbf{x} \ll \beta^{\top}Bu  =\lambda^*\tilde{\varphi}^*(\mathbf{x}) $ leading to the system of approximated Koopman eigenfunctions with eigenvalues along the diagonal of $\lambda^* = diag(\lambda^*_1,\dots,\lambda^*_m)$
\begin{align}
        \dot{\tilde{\varphi}}^{*}(\mathbf{x}) \approx \lambda^*\tilde{\varphi}^*(\mathbf{x})
\end{align}

\subsection{Proof of Stability}
Select the following Lyapunov function candidate
\begin{equation}
    V(x) = \frac{1}{2} \tilde{\varphi}^{{*}^{\top}}{\tilde{\varphi}^{*}} = \mathbf{x}^{\top}\underbrace{\beta\beta^{\top}}_{>0}\mathbf{x} \label{eq:lyapfunc}
\end{equation}
where positive definiteness of $\beta\beta^{\top}$ is guaranteed by $\beta_i$ being composed by Hurwitz polynomial coefficients.
The derivative of \eqref{eq:lyapfunc} is 
\begin{align}
    \dot{V}(\mathbf{x}) &= \tilde{\varphi}^{{*}^{\top}}\left[ \beta^{\top} A \mathbf{x} + \lambda^*\tilde{\varphi}^*(\mathbf{x}) \right] \nonumber\\
    &= \tilde{\varphi}^{{*}^{\top}} \beta^{\top} A \mathbf{x} + \tilde{\varphi}^{{*}^{\top}}\lambda^*\tilde{\varphi}^*(\mathbf{x})  \label{eq:lyapfuncder}
\end{align}
which is negative if condition \eqref{eq:condition1} is followed and all elements of $\lambda^*$ are negative. In fact \eqref{eq:lyapfuncder} is obtained by pre multiplying condition \eqref{eq:condition1} by $\tilde{\varphi}^{{*}^{\top}}$.

\section{Koopman Lower Dimensional Model of Rigid Body Dynamics}\label{sec:Kooplow}
Considering now, the six sets of Koopman generalized eigenfunctions of the rigid body dynamics formulation derived in \cite{martini2024koopman_rigid} and summarized in Section \ref{sec:2c}$, \varkappa = \left( \varkappa_1, \varkappa_2, \varkappa_3, \varkappa_4,\varkappa_5,\varkappa_6 \right)$ where 
\begin{equation}
    \varkappa_{i} = (\{\varkappa_{i,k}\}^{N_{\nu}-1}_{k=0})=(\{\nu_k(j)\}^{N_{\nu}-1}_{k=0})
\end{equation} 
for $i = 1,2,3$ and $j=i$, and 
\begin{equation}
    \varkappa_{i} = (\{\varkappa_{i,k}\}^{N_{z}-1}_{k=0})=(\{z_k(j)\}^{N_{z}-1}_{k=0})
\end{equation}
for $i = 4,5,6$ and $j = i-3$. The dynamics of $\varkappa$ are described by \eqref{eq:completeKsys} which correspond to the class of MIMO systems of \eqref{eq:MIMOsys} but with state dependent control matrix $B_J(\varkappa)$. It is easy to show that the approach of Section \ref{sec:3c} can be used on \eqref{eq:completeKsys} as long as $B_{\mathbf{low}}(\varkappa) = \beta^{\top}B_J(\varkappa)$ is invertible. Assuming $B_{\mathbf{low}}(\varkappa)$ nonsingular we construct the following set of approximated Koopman eigenfunction candidates
\begin{equation}
    \dot{\tilde{\varphi}}^{*}(\varkappa) = \beta^{\top} A_J \varkappa  + \underbrace{\beta^{\top}B_J(\varkappa)}_{B_{\mathbf{low}}(\varkappa)}u
\end{equation}
where $\tilde{\varphi}^{*}(\varkappa) = \tilde{\varphi}^{*}(x) = \beta^{\top}\varkappa=\left[ \tilde{\varphi}^{*}_1,\tilde{\varphi}^{*}_2, \tilde{\varphi}^{*}_3, \tilde{\varphi}^{*}_4, \tilde{\varphi}^{*}_5, \tilde{\varphi}^{*}_6\right]^{\top}$ and
$B_{\mathbf{low}}(\varkappa) = B_{\mathbf{low}}(x) = \left[ \tilde{B}^{*}_1,\tilde{B}^{*}_2,\tilde{B}^{*}_3, \tilde{B}^{*}_4, \tilde{B}^{*}_5,\tilde{B}^{*}_6  \right]^{\mathcal{T}}$. Each scalar approximated Koopman eigenfunction candidate is related to one of the six Jordan blocks related to the six variable to control ($p$ and $\eta$) and the dimension of $\tilde{\varphi}^{*}(\varkappa) \in \mathbb{R}^{6\times 1}$ matches the six control inputs ($\tau,F$).
Satisfying condition \eqref{eq:condition1}, the resulting dynamics is approximated to 
\begin{equation}
        \dot{\tilde{\varphi}}^{*}(\varkappa) \approx \mathbf{0}_6 \cdot\tilde{\varphi}^{*}+B_{\mathbf{low}}(\varkappa)u \label{eq:reducedKsysy}
\end{equation}
which can be stabilized by the control input \eqref{eq:control_phiBlowinv}
\begin{equation}
    u = -B_{\mathbf{low}}^{-1}(x)u^* = -B_{\mathbf{low}}^{-1}(x)k\tilde{\varphi}^*(x) \label{eq:control_phiBlowinvrigid}
\end{equation}
which yield to 
\begin{equation}
    \dot{\tilde{\varphi}}^{*}(x) \approx -k\tilde{\varphi}^*(x)\label{eqn:newKobslinear}
\end{equation}
with ${k} = diag(k_1,k_2,k_3,k_4,k_5,k_6)$ such that $\beta^{\top} A_J \varkappa \ll -k\tilde{\varphi}^*(x)=-k\beta^{\top}\varkappa$.
If the conditions of Section \ref{sec:method} are met, \eqref{eq:control_phiBlowinvrigid} is a stabilizing controller for  \eqref{eq:completeKsys} and therefore \eqref{eq:eqmot} (assuming a good approximation of the Koopman rigid body formulation). Thus, the reduced lower dimensional quasi-linear formulation \eqref{eq:reducedKsysy} embeds enough of the original system nonlinear dynamics to be employed for controller design.

\section{Quadrotor Controller Design Case Study}\label{sec:quadrotor}
We apply the proposed methodology to the underactuated quadrotor UAV
\begin{subequations}
    \begin{align}\label{eq:eqmot1}
\dot R &= RS(\nu) \\ J\dot \nu &= M - S(\nu)J\nu \\ \dot p &= Rv \\ \dot v &= \frac{1}{m}T\mathbf{e_3} - S(\nu)v - gR^{\top}\mathbf{e_3}
\end{align}
\end{subequations}
The only key difference with respect to the general rigid body dynamics is that the quadrotor propeller can only generate thrust $T$ in the body attached $z$-axis, hence $F = T\mathbf{e_3}$. Hence, the system input vector is reduced to $\left[\tau,T\right]^{\mathcal{T}} \in \mathbb{R}^{4}$ instead of $u =\left[\tau,T\mathbf{e_3}\right]^{\mathcal{T}} \in \mathbb{R}^{6}$.
From \cite{martini2024koopman_rigid}, the resulting system achieves a quasi-linear form which does not feature underactuation.
To achieve a sufficiently rich embedding of the nonlinear quadrotor dynamics, system dimension of $N = 16$ is selected. Hence the need to resort to the least square approximation method to design the control law. Despite previous successful results, the associated high dimensionality intensifies the controller tuning process. Moreover, the use of least square approximation increases the approximation error, already caused by the Koopman formulation truncation.

\begin{figure}[h]
    \centering
    \includegraphics[width=1\linewidth]{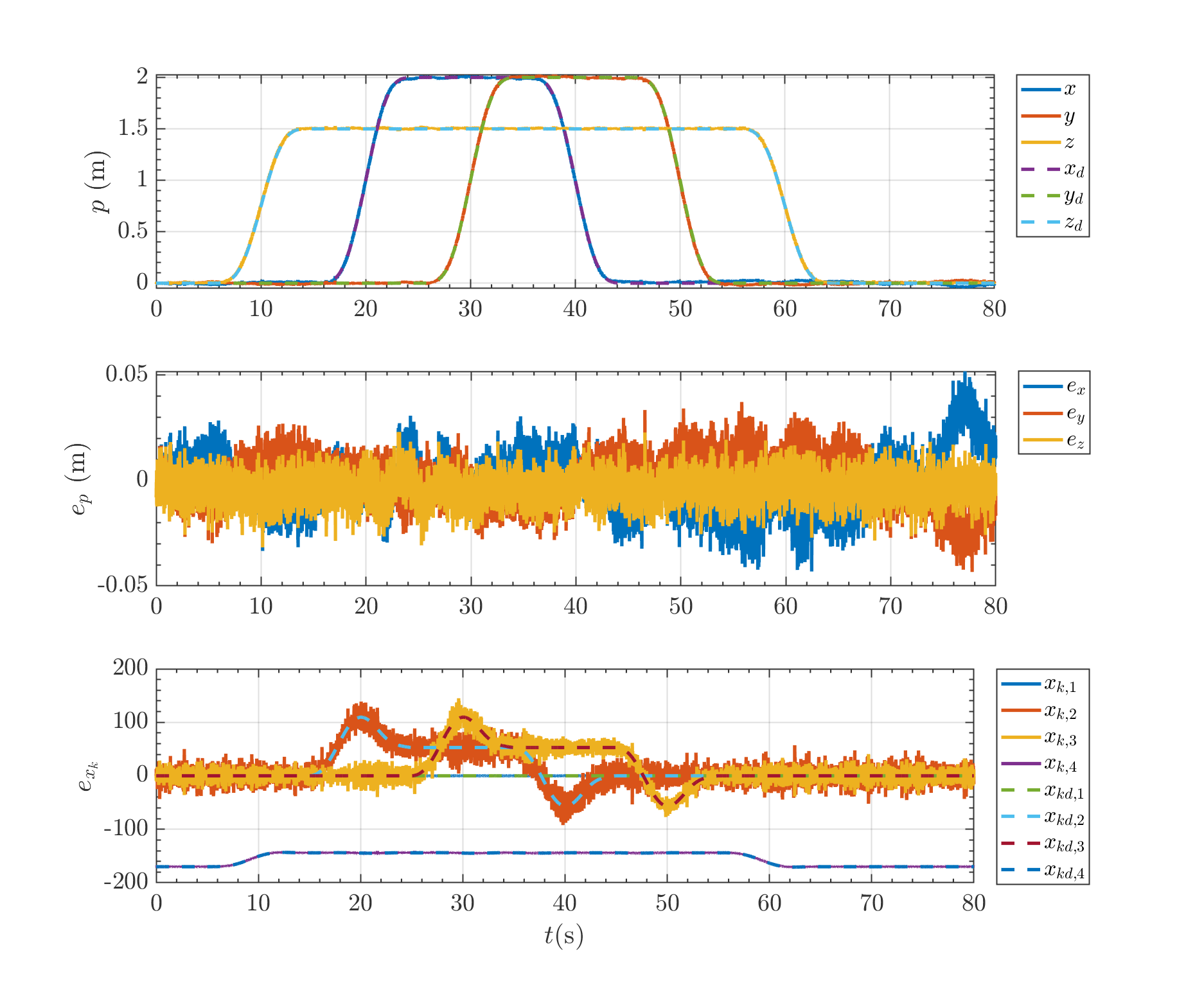}
    \caption{Numerical Simulation Tracking}
    \label{fig:coord}
\end{figure}

\begin{figure}[h]
    \centering
    \includegraphics[width=1\linewidth]{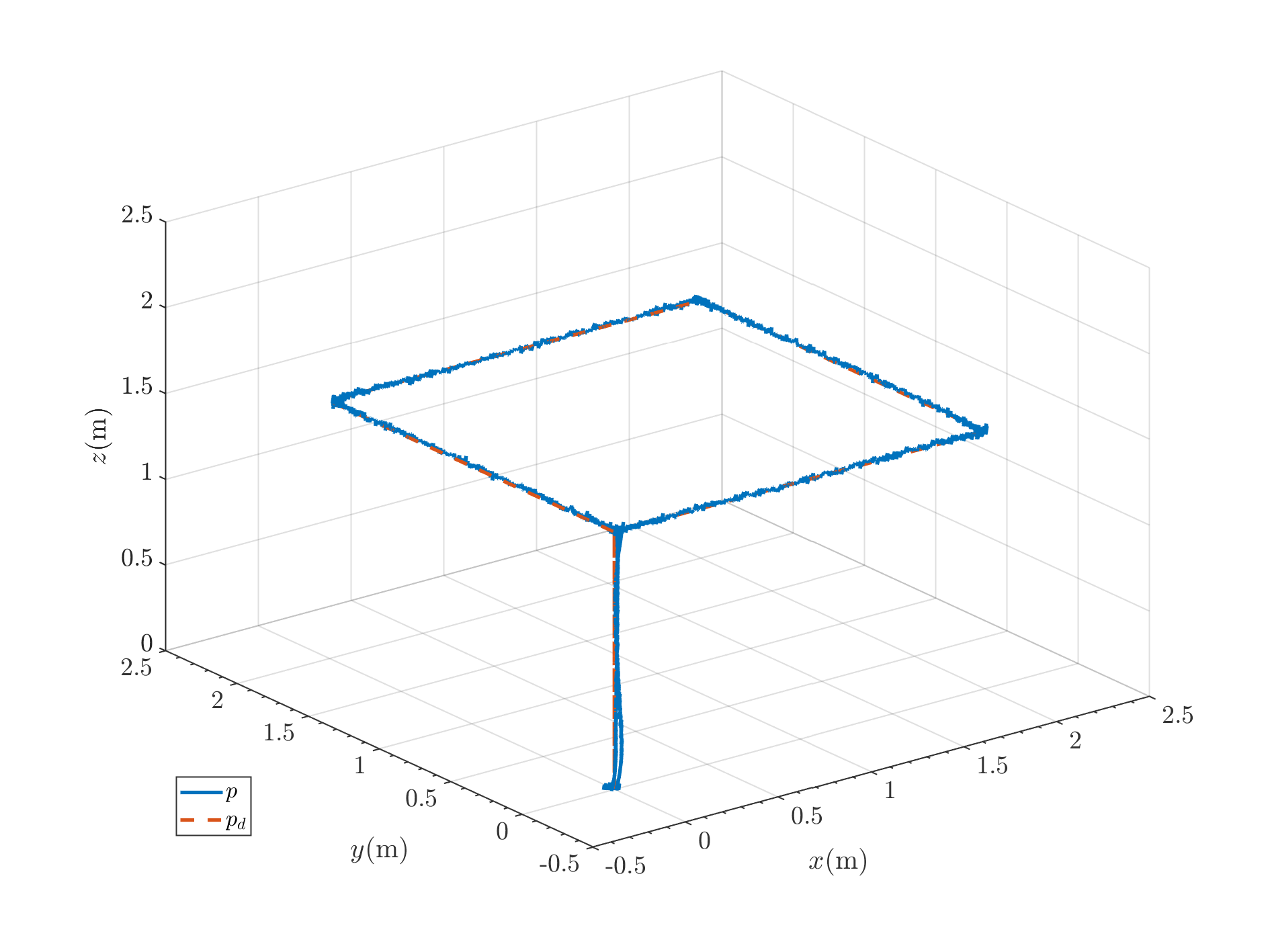}
    \caption{Numerical Simulation 3D Trajectory}
    \label{fig:3D_sim}
\end{figure}

\begin{figure}[h]
    \centering
    \includegraphics[width=1\linewidth]{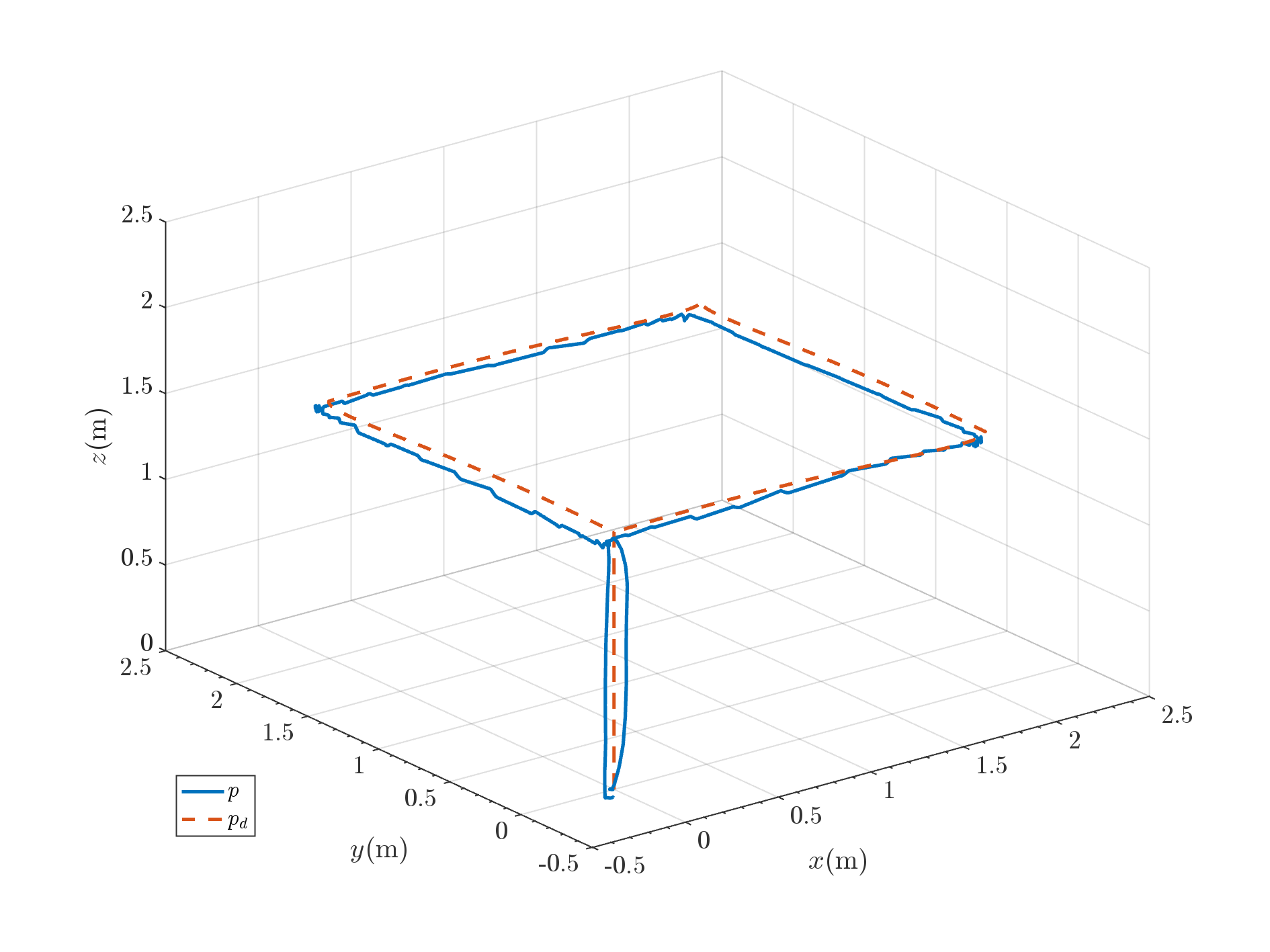}
    \caption{HIL 3D Trajectory}
    \label{fig:3D}
\end{figure}

By applying the proposed methodology to each set of the quadrotor Koopman generalized eigenfunctions we obtain the vector of four approximated Koopman eigenfunction candidates $\Upsilon = \left[ \tilde{\varphi}^{*}_1,\tilde{\varphi}^{*}_4, \tilde{\varphi}^{*}_5, \tilde{\varphi}^{*}_6\right]^{\top}\in \mathbb{R}^{4\times 1}$, each related to one of the four Jordan blocks to control
($\tilde{\varphi}^{*}_2,\tilde{\varphi}^{*}_3$ are discarded since the roll and pitch angular velocity dynamics are embedded within the other generalized eigenfunctions\cite{martini2024koopman_rigid}).
The resulting dynamics are
\begin{equation}
    \dot \Upsilon(x) \approx \mathbf{0}_{4} \cdot \Upsilon(x) + B_{\mathbf{low4}}(x)u_{\mathbf{4}} = u^*\label{eq:newKobslinear}
\end{equation}
where $u_{\mathbf{4}},u^*\in \mathbb{R}^4$ and $B_{\mathbf{low4}}(x) = \left[ \bar{\tilde{B}}^{*}_1, \bar{\tilde{B}}^{*}_4, \bar{\tilde{B}}^{*}_5, \bar{\tilde{B}}^{*}_6  \right]^{\mathcal{T}}\in \mathbb{R}^{4\times 4}$ where $\bar{\tilde{B}}^{*}_i$ represent $\tilde{B}^{*}_i$ where column $4$ and $5$ are removed (since they are not affected by any input $\left[\tau,T\right]^{\mathcal{T}} \in \mathbb{R}^{4}$). Having matched the number of input and output, the same steps of Section \ref{sec:Kooplow} are applied.
\begin{figure*}[h]
    \centering
    \includegraphics[width=0.8\linewidth]{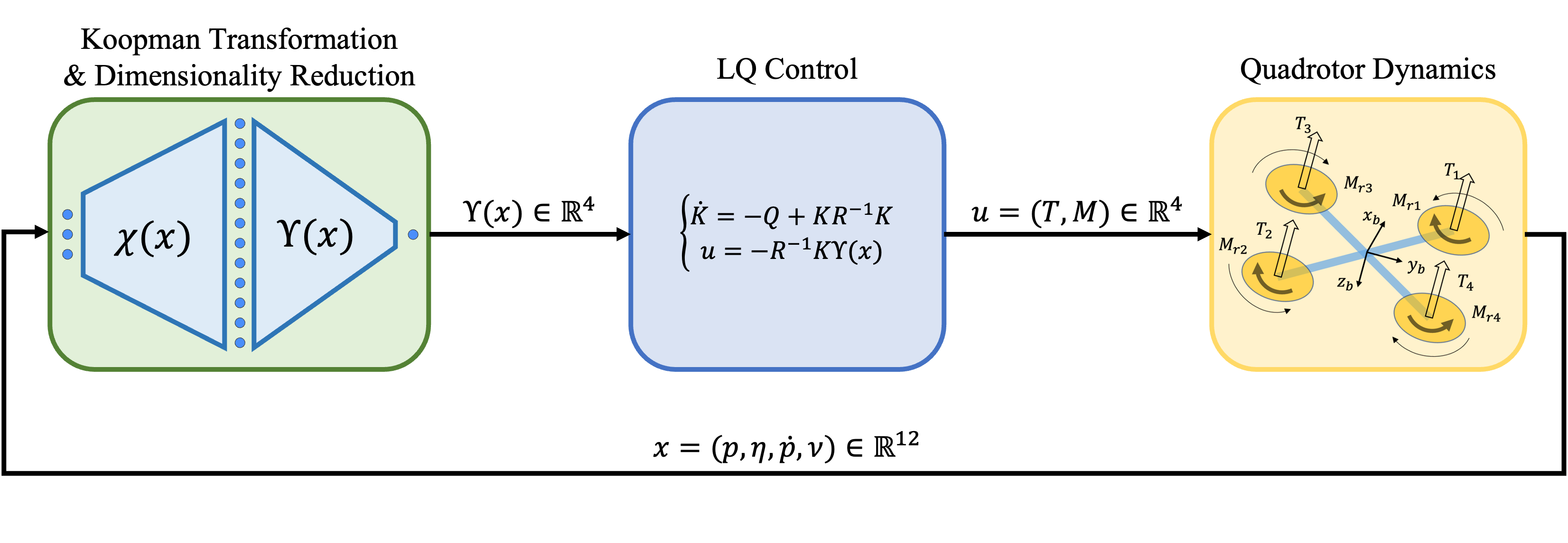}
    \caption{Low Dimensional K-LQ Controller Schematic}
    \label{fig:schematic}
\end{figure*}
Starting from \eqref{eq:newKobslinear} we proceed by designing a LQ controller in terms of the Koopman approximated eigenfunctions according to the schematic in Fig. \ref{fig:schematic}. Define the error $e_{\Upsilon}(x) = \Upsilon_d(x_d) - \Upsilon(x)$ which dynamics are 
\begin{equation}
    \dot e_{\Upsilon}(x) = -u^* + u_d^* = U^* \label{eqn:newKerrorlinear}
\end{equation}
The LQ control is designed to minimize the following cost function
\begin{equation}
    J = \frac{1}{2}\int_0^{t_f}  {({e_{\Upsilon}^{\top}}Q_{LQ}e_{\Upsilon} + U^{*\top}R_{LQ}U^*)dt} 
\end{equation}
where $Q_{LQ}$ and $R_{LQ}$ are weight matrices to be tuned.
It follows that the resulting control input is computed as 
\begin{equation}
    u = B_{\mathbf{low4}}^{-1}(x)\left( R_{LQ}^{-1}Ke_{\Upsilon} + u_d^* \right)
\end{equation}
where $K$ is the solution of the algebric Riccati differential equation
\begin{equation}
    0 = -Q_{LQ} + KR_{LQ}^{-1}K
\end{equation}
The closed loop dynamics are
\begin{equation}
    \dot e_{\Upsilon}(x) = -ke_{\Upsilon}(x) + u_d^*
\end{equation}
where $k = R^{-1}K$.
The resulting control law is implemented for the trajectory tracking task of a square trajectory. The quadrotor parameters align with those presented in \cite{sonmez2025Reinforcement}. $B_{\mathbf{low4}}(x)$ is verified to be nonsingular through symbolic check.  Fig \ref{fig:coord} and \ref{fig:3D_sim} shows that a good tracking of the Koopman approximated eigenfunctions correspond to a desirable tracking performance in the original state space. Moreover, throughout the simulation $\beta^{\top} A_J \varkappa $ remains one order of magnitude smaller than $-k\beta^{\top}\varkappa$.
Finally, hardware in the loop (HIL) experiments are performed.
The K-LQ controller is deployed on pixhawk 2.1 cube black while the quadrotor dynamics is simulated using Jmavsim simulator using a similar setup to \cite{sonmez2025Reinforcement}. 
As shown in figure \ref{fig:3D} the controller is able to guarantee desirable tracking performance highlighting the real-time implementability of the proposed methodology.
\section{Conclusions}\label{sec:conclusions}
In this work, we presented a methodology to achieve a low dimensional Koopman representation based on Koopman generalized eigenfunctions. The novel formulation is designed to match the number of inputs to the number of outputs in order to avoid the least square approximation when performing Koopman based linearization. The obtained formulation, when applied to the rigid body dynamics leads to a quasi-linear Koopman based model, which can be linearized resulting into a single integrator decoupled MIMO dynamics. Practically, any linear controller can be designed in the presented formulation. When applied to the quadrotor case study, the resulting quasi-linear model is fully-actuated and it is used to design a single loop LQ flight controller. The presented hardware in the loop results showcase the methodology potential in real-time applications in presence of noise and feedback delays. The unique results achieved in this work highlight the practical advantages of Koopman based strategies posing them as viable solution to the underactuation problem. Future works will focus on deriving an exact bound of condition \eqref{eq:condition1}.

\appendix
\section{Interpretation of PID Control through the Koopman Operator}\label{appendix}

\large
\textbf{Review: Explain PID with Koopman for Higher order Integrators}\label{part:1}
\normalsize

\subsection{Double Integrator Plant}
Let's consider the class of ODE which are characterized by a double integrator dynamics.
\begin{equation}
    \ddot p = u
\end{equation}
where $p$ is the variable to be controlled, $u$ is the input forcing term. This class of system include a wide selection of linear dynamic processes for example the mass spring dumper system
\begin{equation}
    \ddot p + \frac{c}{m} \dot p + \frac{k}{m} p = \frac{1}{m}u^* \quad \underset{u^* = c \dot p + k p + m u}{\Longrightarrow} \quad \ddot p = u
\end{equation}
and control-affine nonlinear dynamics exactly linearized with dynamics compensation
\begin{equation}
    \ddot p = f(p,\dot p) + gu^* \quad \underset{u^* = g^{-1}(-f(p) + u)}{\Longrightarrow} \quad \ddot p = u
\end{equation}
The state space realization for the double integrator dynamics is commonly written as
\begin{equation}
    \underbrace{\begin{bmatrix} \dot p \\ {\ddot{p}} \end{bmatrix}}_{\dot {\mathbf{x}}} = \underbrace{\begin{bmatrix} 0 & 1 \\ 0 & 0 \end{bmatrix}}_{A} \underbrace{\begin{bmatrix} p \\ \dot p \end{bmatrix}}_{\mathbf{x}} + \underbrace{\begin{bmatrix} 0 \\ 1 \end{bmatrix}}_B u
\end{equation}

\begin{equation}
    \underbrace{\begin{bmatrix} p \\ \dot p \\ {\ddot{p}} \end{bmatrix}}_{\dot {\mathbf{x}}} = \underbrace{\begin{bmatrix} 0 & 1 & 0 \\ 0 & 0 & 1 \\ 0 & 0 & 0 \end{bmatrix}}_{A} \underbrace{\begin{bmatrix} \int{p} \\ p \\ \dot p \end{bmatrix}}_{\mathbf{x}} + \underbrace{\begin{bmatrix} 0 \\ 0 \\ 1 \end{bmatrix}}_B u
\end{equation}

while the transfer function form is defined as
\begin{equation}
    G(s) = \frac{P(s)}{U(s)} = \frac{1}{s^2}
\end{equation}
\subsection{PID Pole Placement}

The PID controller is given by:

\[
C(s) = K_p + \frac{K_i}{s} + K_d s
\]

The plant is:

\[
G(s) = \frac{1}{s^2}
\]

The open-loop transfer function is:

\[
L(s) = C(s) \cdot G(s) = \left(K_p + \frac{K_i}{s} + K_d s\right) \cdot \frac{1}{s^2} = \frac{K_p}{s^2} + \frac{K_i}{s^3} + \frac{K_d}{s}
\]

The closed-loop transfer function with unity feedback is:

\[
T(s) = \frac{L(s)}{1 + L(s)} = \frac{\frac{K_p}{s^2} + \frac{K_i}{s^3} + \frac{K_d}{s}}{1 + \frac{K_p}{s^2} + \frac{K_i}{s^3} + \frac{K_d}{s}}
\]

The characteristic equation is the denominator of the closed-loop transfer function:

\[
1 + L(s) = 1 + \frac{K_p}{s^2} + \frac{K_i}{s^3} + \frac{K_d}{s} = 0
\]

Multiplying through by \( s^3 \) to eliminate the fractions:

\[
s^3 + K_d s^2 + K_p s + K_i = 0
\]

which correspond to the state space system
\begin{equation}
    \underbrace{\begin{bmatrix} p \\ \dot p \\ {\ddot{p}} \end{bmatrix}}_{\dot {\mathbf{x}}} = \underbrace{\begin{bmatrix} 0 & 1 & 0 \\ 0 & 0 & 1 \\ -k_i & -k_p & -k_d \end{bmatrix}}_{A} \underbrace{\begin{bmatrix} \int{p} \\ p \\ \dot p \end{bmatrix}}_{\mathbf{x}} + \underbrace{\begin{bmatrix} 0 \\ 0 \\ 1 \end{bmatrix}}_B u
\end{equation}
We now want to place the poles of the system at desired locations \( \lambda_1 \), \( \lambda_2 \), and \( \lambda_3 \). The desired characteristic polynomial with poles at \( \lambda_1 \), \( \lambda_2 \), and \( \lambda_3 \) is:

\[
(s - \lambda_1)(s - \lambda_2)(s - \lambda_3)
\]

Expanding this product gives the characteristic equation:

\[
s^3 - (\lambda_1 + \lambda_2 + \lambda_3) s^2 + (\lambda_1 \lambda_2 + \lambda_1 \lambda_3 + \lambda_2 \lambda_3) s - \lambda_1 \lambda_2 \lambda_3 = 0
\]

Comparing this with the standard form \( s^3 + K_d s^2 + K_p s + K_i = 0 \), we can match the coefficients:

\[
K_d = -(\lambda_1 + \lambda_2 + \lambda_3)
\]
\[
K_p = \lambda_1 \lambda_2 + \lambda_1 \lambda_3 + \lambda_2 \lambda_3
\]
\[
K_i = - \lambda_1 \lambda_2 \lambda_3
\]

Thus, the controller gains \( K_d \), \( K_p \), and \( K_i \) are directly related to the desired poles \( \lambda_1 \), \( \lambda_2 \), and \( \lambda_3 \).

\subsubsection{Case 1: Three Distinct and Real Negative Eigenvalues}

Let the eigenvalues \( \lambda_1, \lambda_2, \lambda_3 \) be distinct and real negative values, i.e., \( \lambda_1, \lambda_2, \lambda_3 < 0 \).

In this case, the characteristic polynomial becomes:

\[
(s - \lambda_1)(s - \lambda_2)(s - \lambda_3) = s^3 - (\lambda_1 + \lambda_2 + \lambda_3) s^2 + (\lambda_1 \lambda_2 + \lambda_1 \lambda_3 + \lambda_2 \lambda_3) s - \lambda_1 \lambda_2 \lambda_3
\]

Comparing this with the characteristic equation \( s^3 + K_d s^2 + K_p s + K_i = 0 \), we can directly compute the controller gains:

\[
K_d = -(\lambda_1 + \lambda_2 + \lambda_3)
\]
\[
K_p = \lambda_1 \lambda_2 + \lambda_1 \lambda_3 + \lambda_2 \lambda_3
\]
\[
K_i = -\lambda_1 \lambda_2 \lambda_3
\]

Therefore, the PID controller gains are determined by the sum and products of the desired eigenvalues.

\subsubsection{Case 2: One Real Negative Eigenvalue and Two Complex Conjugates with Negative Real Part}

Let \( \lambda_1 \) be a real negative eigenvalue, and \( \lambda_2 \) and \( \lambda_3 \) be complex conjugates with negative real parts. The eigenvalues \( \lambda_2 \) and \( \lambda_3 \) can be written as:

\[
\lambda_2 = \alpha + j\beta, \quad \lambda_3 = \alpha - j\beta
\]

where \( \alpha < 0 \) is the negative real part, and \( j \beta \) is the imaginary part.

The characteristic polynomial is:

\[
(s - \lambda_1)(s - \lambda_2)(s - \lambda_3) = (s - \lambda_1)(s^2 - 2\alpha s + (\alpha^2 + \beta^2))
\]

Expanding this:

\[
(s - \lambda_1)(s^2 - 2\alpha s + (\alpha^2 + \beta^2)) = s^3 - (\lambda_1 + 2\alpha) s^2 + (\lambda_1 \cdot 2\alpha + \alpha^2 + \beta^2) s - \lambda_1 (\alpha^2 + \beta^2)
\]

Now, comparing this with the characteristic equation \( s^3 + K_d s^2 + K_p s + K_i = 0 \), we get the controller gains:

\[
K_d = -(\lambda_1 + 2\alpha)
\]
\[
K_p = \lambda_1 \cdot 2\alpha + (\alpha^2 + \beta^2)
\]
\[
K_i = -\lambda_1 (\alpha^2 + \beta^2)
\]

Thus, for this case, the PID controller gains are determined by the real and imaginary parts of the complex eigenvalues \( \lambda_2 \) and \( \lambda_3 \) as well as the real eigenvalue \( \lambda_1 \).

Let \( \lambda_1 \) be real and negative, and \( \lambda_2, \lambda_3 \) be complex conjugates with negative real part. The coefficients are:

\[
K_d = -(\lambda_1 + 2 \text{Re}(\lambda_2)), \quad K_p = \lambda_1 \cdot 2 \text{Re}(\lambda_2) + |\lambda_2|^2, \quad K_i = -\lambda_1 |\lambda_2|^2
\]

\subsection{Koopman Point of View}
\textbf{Autonomous systems}
\begin{equation}
     \dot{\mathbf{x}} = f(\mathbf{x})
\end{equation}
\begin{equation}
    \mathcal{K}(b(\mathbf{x})) = \dot b(\mathbf{x}) = \nabla_\mathbf{x} b(\mathbf{x})  f(\mathbf{x})
\end{equation}
\begin{equation}
    \dot{\mathbf{x}} = A \mathbf{x}
\end{equation}
\begin{equation}
    \mathcal{K}(\mathbf{x}) = \dot{\mathbf{x}} = \underbrace{\nabla_\mathbf{x} \mathbf{x}}_{I} A\mathbf{x} = A \mathbf{x}
\end{equation}
Hence for linear systems the state matrix $A$ is an infinitesimal generator of the Koopman operator.
\textbf{Forced systems}
\begin{equation}
     \dot{\mathbf{x}} = f(\mathbf{x}) + G(\mathbf{x})u
\end{equation}
\begin{equation}
    \mathcal{K}(b(\mathbf{x})) = \dot b(\mathbf{x}) = \nabla_\mathbf{x} b(\mathbf{x})  f(\mathbf{x}) + \nabla_\mathbf{x} b(\mathbf{x})  G(\mathbf{x})u
\end{equation}
\begin{equation}
    \dot{\mathbf{x}} = A \mathbf{x} + Bu
\end{equation}
\begin{equation}
    \mathcal{K}(\mathbf{x}) = \dot{\mathbf{x}} = \underbrace{\nabla_\mathbf{x} \mathbf{x}}_{I} (A\mathbf{x}  + Bu)= A \mathbf{x} + Bu
\end{equation}

\subsubsection{Transfer Function}

\begin{align}
    &s(s - \lambda_1)(s - \lambda_2) = \lambda_0(s - \lambda_1)(s - \lambda_2)\rightarrow s\left( s^2 - (\lambda_1 + \lambda_2)s + \lambda_1 \lambda_2 \right) = \lambda_0 \left( s^2 - (\lambda_1 + \lambda_2)s + \lambda_1 \lambda_2 \right) \\
    &\rightarrow s\left( s - (\lambda_1 + \lambda_2) + \lambda_1 \lambda_2\frac{1}{s} \right) = \lambda_0 \left( s - (\lambda_1 + \lambda_2) + \lambda_1 \lambda_2 \frac{1}{s}\right)\\
    &\overset{\mathcal{L}}{\rightarrow} \frac{d}{dt}{\left( \dot{p} + - (\lambda_1 + \lambda_2)p +  \lambda_1 \lambda_2 \int{p}  \right)} =  \lambda_0 \left( \dot{p} + - (\lambda_1 + \lambda_2)p +  \lambda_1 \lambda_2 \int{p}  \right)
\end{align}
Hence we see that the desired characteristic polynomial leads to a set of three ODEs which in terms of Koopman eigenfunctions
\begin{align}
    \frac{d}{dt}\underbrace{{\left( \dot{p} - (\lambda_1 + \lambda_2)p +  \lambda_1 \lambda_2 \int{p}  \right)}}_{\varphi_0} &=  \lambda_0 \left( \dot{p} - (\lambda_1 + \lambda_2)p +  \lambda_1 \lambda_2 \int{p}  \right)\\
    \frac{d}{dt}\underbrace{{\left( \dot{p} - (\lambda_0 + \lambda_2)p +  \lambda_0 \lambda_2 \int{p}  \right)}}_{\varphi_1} &=  \lambda_1 \left( \dot{p} - (\lambda_0 + \lambda_2)p +  \lambda_0 \lambda_2 \int{p}  \right)\\
    \frac{d}{dt}\underbrace{{\left( \dot{p} - (\lambda_0 + \lambda_1)p +  \lambda_0 \lambda_1 \int{p}  \right)}}_{\varphi_2} &=  \lambda_2 \left( \dot{p} - (\lambda_0 + \lambda_1)p +  \lambda_0 \lambda_1 \int{p}  \right)
\end{align}

\begin{align}
    \frac{d}{dt}\underbrace{{\left( \dot{p} - (\lambda_1 + \lambda_2)p +  \lambda_1 \lambda_2 \int{p}  \right)}}_{\varphi_0} &=  - (\lambda_1 + \lambda_2) \dot p +  \lambda_1 \lambda_2 p  + u \\
    \frac{d}{dt}\underbrace{{\left( \dot{p} - (\lambda_0 + \lambda_2)p +  \lambda_0 \lambda_2 \int{p}  \right)}}_{\varphi_1} &=  - (\lambda_0 + \lambda_2) \dot p +  \lambda_0 \lambda_2 p  + u \\
    \frac{d}{dt}\underbrace{{\left( \dot{p} - (\lambda_0 + \lambda_1)p +  \lambda_0 \lambda_1 \int{p}  \right)}}_{\varphi_2} &=  - (\lambda_0 + \lambda_1) \dot p +  \lambda_0 \lambda_1 p  + u 
\end{align}

\begin{equation}
    u = -k_i \int p -k_p p -k_d \dot p
\end{equation}
We can force the control parameters so that 

\begin{align}
 \lambda_0 \left( \dot{p} - (\lambda_1 + \lambda_2)p +  \lambda_1 \lambda_2 \int{p}  \right) = - (\lambda_1 + \lambda_2) \dot p +  \lambda_1 \lambda_2 p  -k_i \int p -k_p p -k_d \dot p \\
  \lambda_1 \left( \dot{p} - (\lambda_0 + \lambda_2)p +  \lambda_0 \lambda_2 \int{p}  \right) = - (\lambda_0 + \lambda_2) \dot p +  \lambda_0 \lambda_2 p  -k_i \int p -k_p p -k_d \dot p \\
   \lambda_2 \left( \dot{p} - (\lambda_0 + \lambda_1)p +  \lambda_0 \lambda_1 \int{p}  \right) = - (\lambda_0 + \lambda_1) \dot p +  \lambda_0 \lambda_1 p  -k_i \int p -k_p p -k_d \dot p \\
\end{align}
which is achieved by setting
\begin{align}
K_d &= -( \lambda_0 +\lambda_1 + \lambda_2 )\\
K_p &= \lambda_1 \lambda_2 + \lambda_0 \lambda_2 + \lambda_0 \lambda_2\\
K_i &= -\lambda_0 \lambda_1 \lambda_2
\end{align}
which lead to the same result as the pole placement.

Hence, given a set of desired closed loop eigenvalues, a PID controller tuning is one that makes a linear combination of the open loop Koopman generalized eigenfunctions into closed loop Koopman Operator eigenfunctions. Interestingly enough, the pole placement can be achieved by writing only one of the Koopman eigenfunctions since the remaining ones are strictly dependent and uniquely defined.

The resulting systems can be rewritten as
\begin{equation}
    \underbrace{\dot{\begin{bmatrix}\varphi_0 \\ \varphi_1 \\ \varphi_2 \end{bmatrix}}}_{\dot \varphi} = \underbrace{\begin{bmatrix} \varphi_0 & 0 & 0 \\ 0 & \varphi_1 & 0 \\ 0 & 0 & \varphi_2 \end{bmatrix}}_{\Lambda} \underbrace{{\begin{bmatrix} \varphi_0 \\ \varphi_1 \\ \varphi_2 \end{bmatrix}}}_{\varphi}
\end{equation}

\subsubsection{State Space}
\begin{equation}
    \dot{\mathbf{x}} = (A - BK)\mathbf{x}
\end{equation}
\begin{align}
    &\mathcal{K}(\varphi(\mathbf{x})) = \dot\varphi({\mathbf{x}}) = \underbrace{\nabla_\mathbf{x} \varphi({\mathbf{x}})}_{V} (A - BK)\mathbf{x}= \Lambda \varphi({\mathbf{x}}) \\
    &\rightarrow V(A-BK)V^{-1}\underbrace{V\mathbf{x}}_{\varphi(\mathbf{x})} = \Lambda \varphi(\mathbf{x})\\
    &\rightarrow V(A-BK)V^{-1} = \Lambda \\
    &\rightarrow BK = A- V^{-1}\Lambda V\\
    & \rightarrow K = B^{\dagger}(A- V^{-1}\Lambda V)
\end{align}
where $V$ is the matrix of closed loop eigenvectors
\begin{equation}
    V = \begin{bmatrix}
        \lambda_1 \lambda_2 & - (\lambda_1 + \lambda_2) & 1\\
        \lambda_0 \lambda_2 & - (\lambda_0 + \lambda_2) & 1\\
        \lambda_0 \lambda_1 & - (\lambda_0 + \lambda_1) & 1
    \end{bmatrix}
\end{equation}
Hence pole placement is achieved by diagonalizing the closed loop system.

The proposed methodology can be extended to higher order integrators, however, the control resulting controller is no more a PID scheme.

Consider the higher order integrator system
\begin{equation}
    p^{(n)} = u
\end{equation}
which transfer function is 
\begin{equation}
    G(s) = \frac{1}{s^n}
\end{equation}
and state space representation is
\begin{equation}
    \dot{\mathbf{x}} = A \mathbf{x} + Bu
\end{equation}
where
\begin{align}
    \mathbf{x} = \begin{bmatrix}
        \int p \\ p^{(0)} \\ \vdots \\ p^{(n)}
    \end{bmatrix},
    \quad A = \begin{bmatrix}
        0 & 1  & \cdots & 0\\
        0 & 0 & \ddots & \vdots\\
        0 & 0 & \cdots & 1\\
        0 & 0 & \cdots & 0\\
    \end{bmatrix},
    \quad B = \begin{bmatrix}
        0 \\ \vdots \\ 0 \\ 1
    \end{bmatrix}
\end{align}
After proper selection of desired closed loop eigenvalues $\lambda_0, \dots, \lambda_{n+1}$, the higher order integrator system can be controlled by
$u = K\mathbf{x}$
where
\begin{equation}\label{eq:Kother}
    K = B^{\dagger}(A- V^{-1}\Lambda V)
\end{equation}
where the $i$-th line of $V$ is composed by the coefficient of the polynomial $\prod\limits_{k=0}^{n} s-\lambda_{k}$ for all $k\neq i$, in flipped order.
The described approach is generalized for any vector $\mathbf{x}$ 
\begin{equation}
     \mathbf{x} =  \begin{bmatrix} \mathbf{x_0} & \mathbf{x_1} & \dots & \mathbf{x_{n+1}} \end{bmatrix}^{\top}
\end{equation}
The described methodology only applies for $\lambda_i \neq 0$ distinct, otherwise the matrix $V$ would lose rank and become singular. If these conditions are met, a unique tuning of $K$ can be computed such that the closed loop system $\dot{\mathbf{x}} = (A-BK) \mathbf{x}$ has eigenvalues $\lambda_i$, hence full pole placement can be achieved.




\large
\textbf{Explain PID with Koopman for General Systems in Jordan Form}
\normalsize

\subsection{General Systems in Jordan Form}
A general high dimensional linear system in Jordan form can be written as 
\begin{equation}
    \dot{\mathbf{x}} = A \mathbf{x} + Bu
\end{equation}
with
\begin{align}
    \mathbf{x} = \begin{bmatrix}
        \mathbf{x_0} \\ \mathbf{x_1} \\ \vdots \\ \mathbf{x_{{n-1}}}
    \end{bmatrix},
    \quad A = \begin{bmatrix}
        0 & 1  & \cdots & 0\\
        0 & 0 & \ddots & \vdots\\
        0 & 0 & \cdots & 1\\
        0 & 0 & \cdots & 0\\
    \end{bmatrix},
    \quad B = \begin{bmatrix}
        b_0 \\ b_1 \\ \vdots \\ b_{n-1}
    \end{bmatrix}
\end{align}
where $\mathbf{x} \in \mathbb{R}^{n\times1}$, $u \in \mathbb{R}^{m\times1}$, $A \in \mathbb{R}^{n\times n}$, $B \in \mathbb{R}^{n\times m}$, $b_i \in \mathbb{R}^{1\times m}$.
\begin{itemize}
    \item Case $m = 1$. Using the method presented in Part \ref{part:1}, does not lead to a closed loop system with desired eigenvalues $\lambda_i$
\end{itemize}

\begin{align}
 \lambda_0 \left( \mathbf{x_2} - (\lambda_1 + \lambda_2)\mathbf{x_1} +  \lambda_1 \lambda_2 \mathbf{x_0}  \right) = - (\lambda_1 + \lambda_2) \mathbf{x_2} +  \lambda_1 \lambda_2 \mathbf{x_1}  + \left(b_2 - (\lambda_1 + \lambda_2)b_1 + \lambda_1 \lambda_2 b_0 \right)(-k_i \mathbf{x_0} -k_p \mathbf{x_1} -k_d \mathbf{x_2}) \label{eq:poleplace1} \\
  \lambda_1 \left( \mathbf{x_2} - (\lambda_0 + \lambda_2)\mathbf{x_1} +  \lambda_0 \lambda_2 \mathbf{x_0}  \right) = - (\lambda_0 + \lambda_2) \mathbf{x_2} +  \lambda_0 \lambda_2 \mathbf{x_1}  +\left(b_2 - (\lambda_0 + \lambda_2)b_1 + \lambda_0 \lambda_2 b_0 \right)(-k_i \mathbf{x_0} -k_p \mathbf{x_1} -k_d \mathbf{x_2}) \label{eq:poleplace2} \\
   \lambda_2 \left( \mathbf{x_2} - (\lambda_0 + \lambda_1)\mathbf{x_1} +  \lambda_0 \lambda_1 \mathbf{x_0}  \right) = - (\lambda_0 + \lambda_1) \mathbf{x_2} +  \lambda_0 \lambda_1 \mathbf{x_1}  +\left(b_2 - (\lambda_0 + \lambda_1)b_1 + \lambda_0 \lambda_1 b_0 \right)(-k_i \mathbf{x_0} -k_p \mathbf{x_1} -k_d \mathbf{x_2}) \label{eq:poleplace3}
\end{align}
which, in the case of $b_0,b_1 \neq 0$, is only feasible for $\lambda_0 = \lambda_1 = \lambda_2 = \lambda$. However when this condition is met, $V$ is rank deficient and so $K$ cannot be computed with \eqref{eq:Kother}. One possibility is to rewrite \eqref{eq:poleplace1},\eqref{eq:poleplace2} and \eqref{eq:poleplace3} as 
\begin{equation}
     \lambda \left( \mathbf{x_2} - 2\lambda \mathbf{x_1} +  \lambda^2 \mathbf{x_0}  \right) = - 2\lambda \mathbf{x_2} +  \lambda^2 \mathbf{x_1}  + \underbrace{\left(b_2 - 2\lambda b_1 + \lambda^2 b_0 \right)}_{b}(-k_i \mathbf{x_0} -k_p \mathbf{x_1} -k_d \mathbf{x_2}) \label{eq:poleplaceLambdaeq}
\end{equation}

the control gains can be found so that \eqref{eq:poleplaceLambdaeq} is a Koopman eigenfunction. However contrary to the result of Part \ref{part:1}, only one, and not three, of the eigenvalue of $A-BK$ is in $\lambda$, limiting the ability to place the desired poles. 
Alternatively we can consider $\lambda_1 = \lambda_2 = \lambda$ and $\lambda_0 = \lambda^*$ leading to
\begin{equation}
     \lambda^* \left( \mathbf{x_2} - 2\lambda \mathbf{x_1} +  \lambda^2 \mathbf{x_0}  \right) = - 2\lambda \mathbf{x_2} +  \lambda^2 \mathbf{x_1}  +  \underbrace{\left(b_2 - 2\lambda b_1 + \lambda^2 b_0 \right)}_{b^*}(-k_i \mathbf{x_0} -k_p \mathbf{x_1} -k_d \mathbf{x_2}) \label{eq:poleplaceLambdaeq}
\end{equation}

\begin{equation}
    \lambda^* \left( \beta_2\mathbf{x_2} +\beta_1\mathbf{x_1} +  \beta_0 \mathbf{x_0}  \right) = \beta_1 \mathbf{x_2} +  \beta_0 \mathbf{x_1}  + \underbrace{\left(\beta_2 b_2 +\beta_1 b_1 + \beta_0 b_0 \right)}_{b^*}(-k_0 \mathbf{x_0} -k_1 \mathbf{x_1} -k_2 \mathbf{x_2}) \label{eq:poleplace1}
\end{equation}
where $\beta_0 = \lambda_1 \lambda_2$, $\beta_1 = - (\lambda_1 + \lambda_2)$, and $\beta_2 = 1$ and for which the controller gains can be found as 
\begin{align}
k_0 &= \frac{-\lambda^* \beta_0}{b^*}\\
k_1 &= \frac{-\lambda^* \beta_1 +\beta_0}{b^*}\\
k_2 &= \frac{-\lambda^* \beta_2 +\beta_1}{b^*}
\end{align}
or
\begin{equation}
    K = \left[k_0, k_1, k_2 \right] = \frac{-\lambda^*\beta^{\top} + \beta^{\top}A}{\beta^{\top}B}
\end{equation}
it is clear that 
The remaining two eigenvalues however are still uniquely defined as the solutions of the following polynomial
\begin{equation}
    \alpha_3 \text{z}^2 + \alpha_2 \text{z} +\alpha_1 = 0 \label{eq:eigPolySol}
\end{equation}
where the coefficient can be found as
\begin{equation}
    \alpha_i = \frac{\nabla_x \mathscr{L}^{-1} \{(s-\lambda)^{n-1} \} A^{n-i}B}{\nabla_x \mathscr{L}^{-1} \{(s-\lambda)^{n-1} \} B} \label{eq:eigPolySolCoeff}
\end{equation}
which does not guarantee the remaining eigenvalues to be found in the left hand plane. The coefficient formula is derived from the symbolic solver in MATLAB which exploit Schur decomposition to compute the eigenvalues. This result is preserved for any case of $m<n$ and shows that only one eigenvalue can be placed while the remaining ones depend on the control matrix $B$ and the eigenvalue $\lambda$. On further inspection it can be shown that the \eqref{eq:eigPolySol} and \eqref{eq:eigPolySolCoeff} is the general formula to find the eigenvalues of the closed loop system $A-BK$ where the value of $K$ have been computed using the pole placing method of Part \ref{part:1}. 

The result of Part \ref{part:1} are just a special case for which $B = [0, 0, 1]^{\top}$ which makes the solution of \eqref{eq:eigPolySol} coincide with the selected eigenvalues.







\end{document}